\documentclass[aps,pre,groupedaddress]{revtex4-1}

\usepackage[latin1]{inputenc}
\usepackage{amsmath}
\usepackage{amsfonts}
\usepackage{amssymb}
\usepackage{enumerate}
\usepackage{pst-grad}
\usepackage{natbib}

\newcommand{\opHf}{\mathbf{H}_\omega}
\newcommand{\opHn}{\mathbf{H}_0}
\newcommand{\opHb}{\mathbf{H}}
\newcommand{\opV}{\mathbf{V}_\omega}
\newcommand{\fncV}{v_\omega}
\newcommand{\sysf}{\mathit{S}}
\newcommand{\sysn}{\mathit{S}_0}
\newcommand{\sysV}{\mathit{S}_\omega}
\newcommand{\bath}{\mathit{B}}

\newcommand{\spacen}{\mathcal{H}}
\newcommand{\gibbstate}{\sigma_\beta}
\newcommand{\tr}{\text{Tr}}
\newcommand{\aveb}[1]{\langle #1 \rangle}

\newcommand{\avef}[1]{\langle #1 \rangle_{\beta}}

\newcommand{\Ex}[1]{\mathbb{E}_\omega \left( #1 \right)}
\newcommand{\opVb}{\mathbf{V}}

\newcommand{\conV}{\mathbf{V}}
\newcommand{\equi}{\sigma'_{\beta}}
\newcommand{\opA}{\mathbf{A}}
\newcommand{\x}{\mathbf{x}}

\newcommand{\opUf}{\mathbf{U}_\omega}
\newcommand{\opUb}{\mathbf{U}}
\newcommand{\mapUf}{\mathcal{U}_\omega}
\newcommand{\Lf}{\mathcal{L}_\omega}
\newcommand{\Lb}{\mathcal{L}}
\newcommand{\T}{\mathcal{T}}

\newcommand{\covdiff}[3]{\mathcal{C}_{#1}\left(#2, #3\right)}

\newcommand{\LI}{\mathcal{L}_I}
\newcommand{\opB}{\mathbf{B}}
\newcommand{\h}{h_\opB}

\newcommand{\resp}[2]{\chi_{#1 #2}}
\newcommand{\corr}[2]{C_{#1 #2}}

\newcommand{\hresp}[2]{\hat{\chi}_{#1 #2}}
\newcommand{\hcorr}[2]{\hat{C}_{#1 #2}}

\begin{document}

\author{Fattah  Sakuldee}
\email[Corresponding Author: ]{fattah.sak@student.mahidol.ac.th}
\author{Sujin Suwanna}
\affiliation{MU-NECTEC Collaborative Research Unit on Quantum Information, Department of Physics, Faculty of Science, Mahidol University, Bangkok, Thailand, 10400.}

\title{Effects from Additional Random Configuration to Linear Response and Modified Fluctuation-Dissipation Relation}
\date{\today}
\pacs{}
\begin{abstract}
	In this work, a physical system described by Hamiltonian $\opHf = \opHn + \opV(\x,t)$ consisted of a solvable model $\opHn$ and external random and time-dependent potential $\opV(\x,t)$ is investigated. Under the conditions that the average external potential with respect to the configuration $\omega$ is constant in time, and, for each configuration, the potential changes smoothly that the evolution of the system follows Schr\"odinger dynamics, the mean-dynamics can be derived from taking average of the equation with respect to configuration parameter $\omega$. It provides extra contributions from the deviations of the Hamiltonian and evolved state along the time to the Heisenberg and Liouville-von Neumann equations. Consequently, the Kubo's formula and the fluctuation-dissipation relation obtained from the construction is modified in the sense that the contribution from the information of randomness and memory effect from time-dependence are present.
\end{abstract}

\maketitle

\section{Introduction}
The fluctuation-dissipation relation (FDR) is one of the most well-known formulae in statistical physics especially in weakly perturbed systems around equilibria. The relation was first coined by H. B. Callen and T. A. Welton in 1951 \cite{1951CallenWelton} and developed by R. Kubo in 1957 \cite{1966Kubo}. It states that the rate of energy dissipation, which one can obtain from a measurement, can be described via a fluctuation of the system in terms of the correlation amongst the group of considered quantities \cite{1951CallenWelton,2003Crisanti,2011Cugliandolo,1966Kubo}. It is applicable to explain many physical systems, especially in a measurement process, since one can perturb the considered system to obtain a response function and infer  properties of the system.

However, reports over the decades have indicated that the FDR does not hold appropriately in many situations such as glassy or driven systems. In a system with very slow change toward equilibrium, which can be due to the composition of many complicate effects concerning additional interactions, the FDR could be modified into Quasi-FDR that the correlation and response functions are extended into a more general form \cite{2003Crisanti,PhysRevLett.78.350}. Such extension covers, both theoretically and experimentally, many cases of models which concern, for instances, the long-time relaxation, structural glasses \cite{PhysRevLett.83.5038,PhysRevLett.84.4473}, spin glasses \cite{PhysRevLett.71.173,Bellon2001,Cuglian1999,1998Enzo,PhysRevLett.106.150603,PhysRevLett.98.097203}, Ising model with dipolar interactions \cite{PhysRevB.60.3013}, spin-boson model \cite{PhysRevB.72.121102}, and the Glauber-Ising chain \cite{Verley2011}. Also, a driven system is another type of Hamiltonian model which is distracted from the formal FDR by a different reason: time dependence of the Hamiltonian. Nonetheless, such a system provides similar characters of dynamics as the glassy class. That is, the significant slowness of evolution toward equilibrium or toward steady state. In essence the time dependent Hamiltonian effect can arise from the memory content of the dynamics which prevents the system to lie in or near the equilibrium component \cite{PhysRevE.71.020101,PhysRevLett.104.220601,PhysRevB.81.041301,chen2008,mauri2006,Lobaskin2006}.

Indeed, in literature, the derivations of such deviation of FDR and Kubo's formula can be obtain by modifying the standard approach involved with perturbation from a bath called linear response formalism \cite{1966Kubo}. There are many suggestions of the modifications presented which the readers can consult Ref.\cite{2011Cugliandolo,Baiesi2009,Baiesi2010,Baiesi2013,Seifert2010} and references therein for further information. Almost all of them gave similar revisions of the modified-FDR; namely, the standard one but added by an extra term into the equation. In this work, we introduce an alternative explanation to obtain a modified Kubo's formula and the modified-FDR by using the role of random, time-dependent potential into consideration. Also, the discussions on characteristics of the model and related topics will be given.

\section{Formulation}
\subsection{A Model}

We consider a system described by a time-dependent Hamiltonian 
	\begin{equation}
		\opHf(\x,t) = \opHn + \opV(\x,t)
	\end{equation}
(denoted by $\opHf(t)$ or $\opHf$ for short), where $\opHn$ is a fixed Hamiltonian, and $\opV(\x,t)$ is multiplicative operator by a random smooth function $\fncV(\x,t)$ of $t$; that is $\opV(\x,t)\varphi = \fncV(\x,t)\varphi$ for each $\varphi$ in the Hilbert space $\spacen$. Here the parameter $\omega$ denotes a configuration in a probability space $(\Omega,\mu)$, with a measure $\mu$ is a measure associated with the configurations $\omega$. It can be assumed that $\opHn$ is solvable so that one can formally describe all properties of the system $\sysn$. $\opV(\x,t)$ describes effects from an external system $\sysV$ connecting to the system $\sysn$, but $\sysV$ process random fluctuation. The role of $\sysV$ is to supply or extract energy by random amount of $\fncV(\x,t)$ at time $t$. This causes the measured total energy of the composite system, denoted by $\sysf=\sysn\cup\sysV$, to depend on time $t$. In our consideration, the composite system of the previous systems is also weakly contacted with a reservoir $\bath$
 with a bath parameter $\beta=1/k_BT$, where $T$ is a bath temperature to make it more realistic in thermodynamic sense, and more convenient to calculate statistical physical quantities.  Remark that a glassy-type system or a driven system can be described in this framework by assigning a specific random profile or time-dependence profile to the potential $\opV(t)$.
%
	
Suppose $\{\phi_k\}$ is a set of eigenfunctions of $\opHn$ in the Hilbert space $\spacen$ corresponding to eigenvalues $\{E_k\}$, where $k = 1,2,\ldots,d_0:=\dim\spacen$. We will restrict our consideration to only the Hilbert space $\spacen$. However, one can also explain the composite system $\sysf$ by another Hilbert space which is possibly larger or smaller than $\spacen$ because of the additional term of Hamiltonian $\opV(\x,t)$. Note that the interaction with the bath is sufficiently weak that it does not change the structure of the composite system, or, in our case, the system $\sysn$ (and $\spacen$). For the latter, we consider the case that the Hamiltonian $\opHn$ is that of a $d_0$-level system. For such a system, one can obtain an equilibrium state or Gibbs state which maximizes the von Neumann entropy among all states with a given energy $ E_0$ \cite{1967Katz,RevModPhys.50.221} as
	\begin{equation}
		\displaystyle\gibbstate := \dfrac{e^{-\beta\opHn}}{\tr \left[e^{-\beta\opHn}\right]} = \sum_{k=1}^{d_0}\left(\dfrac{e^{-\beta E_k}}{\sum_{j=1}^{d_0}e^{-\beta E_j}}\right)\vert\phi_k\rangle\langle\phi_k\vert
		\label{eq:gibbs}
	\end{equation}
which is equilibrium for only the case $\fncV(\x,t) = 0$ at all time $t$. In this case, $E_0=\aveb{\opHb}_{\gibbstate}:=\tr\left[\gibbstate\opHb\right]$, where $\aveb{\opA}_\rho =\tr{(\rho\opA)}$ is an average over a certain state $\rho$ of the operator $\opA$. On the contrary, for the system governed by $\opHf$, it becomes more complicate to find an equilibrium state from a given energy because the Hamiltonian $\opHf$ is time-dependent; hence the energy $\tr\left(\gibbstate\opHf(t)\right)$ of the system in the state $\gibbstate$ is no longer a constant of motion. In this scenario, for a path $\omega$, the state of the form $\dfrac{e^{-\beta\opHf}}{\tr e^{-\beta\opHf}}$ is not an equilibrium state. Moreover, since $\opHf$ also depends on configuration $\omega$, one may not obtain the same result from the different preparations or measurements. Therefore, we will impose additional assumptions to define an appropriate constant of motion to develop another version of equilibrium state for overall evolution including the effect of anomalous potential $\opV(\x,t)$.

\subsection{Equilibrium State}
In order to construct an equilibrium state, we add additional assumptions to the external system $\sysV$. For each configuration $\omega \in \Omega$, $\fncV(\x,t)$ is supposed to be a smooth function of $t$. We denote by $\Ex{X_\omega} := \int_\Omega X_\omega d\mu(\omega)$ the expectation or $\omega$-average of random variable $X_\omega$ with respect to the measure $\mu$ on the probability space $\left(\Omega,\mu \right)$. For simplicity, we assume that: 
	\begin{equation}
		\opVb := \Ex{\opV(\x,t)}  \text{ is a constant (multiplicative operator) in time.} 			
		\label{cond:constV}
	\end{equation}
From the condition \eqref{cond:constV}, we find that $\opHn$ and $\opVb$ are time-independent and are also the constants of motion. Thus, an equilibrium state in this case, called \emph{adjusted equilibrium state}, can be written as \cite{1967Katz}
	\begin{equation}
		\equi = \dfrac{e^{-\beta\left(\opHn + \nu\conV\right)}}{\tr\left({e^{-\beta\left(\opHn + \nu\conV\right)}}\right)},
		\label{eq:equilibrium}
	\end{equation}
	where $\beta$ and $\nu$ are Lagrange multipliers arisen from the formulation in Ref.\cite{1967Katz}. Without loss of generality, we set $\nu=1$ by absorbing the scaling into the random part of Hamiltonian. Note that the parameter $\beta$ must equal with the inverse temperature of the bath when $\opV(\x,t)=0$ for all time $t$.
Moreover, by employing the equivalence between the Schr\"odinger and Heisenberg pictures, for an evolution of the operator $\opA\mapsto\opA(t)$ and its dual evolution for a state $\rho\mapsto\rho(t)$, one obtains
	\begin{equation}
		\label{eq:Schro-Heiequivalence}	\aveb{\opA}_{\rho(t)}=\tr{\left(\rho(t)\opA\right)}=\tr{\left(\rho\opA(t)\right)}=\aveb{\opA(t)}_{\rho}.
	\end{equation}
Exact forms of the evolution maps will be discussed in the following subsection.


\subsection{Dynamics of Operators and States}
For a fixed configuration $\omega$, the dynamics of the system  in this specific realization is well defined formally since $\fncV(\x,t)$ is assumed to be a smooth function. 
The evolution of a self adjoint operator observable $\opA$ and its dual dynamics follow
	\begin{eqnarray}
		\dfrac{d}{dt}\opA_\omega(t)	&=&	i\Lf^t(\opA_\omega(t)) := i\left[\opHf,\opA_\omega(t) \right];\label{eq:operatorevol}\\
		\dfrac{d}{dt}\rho_\omega(t)	&=&	-i\Lf^t(\rho_\omega(t)) = -i\left[\opHf,\rho_\omega(t) \right].\label{eq:stateevol}
	\end{eqnarray}
Furthermore, one can also define the dynamics generated by a \emph{mean} Hamiltonian  $\opHb = \opHn + \opVb$. The mean dynamics of operator $\opA$ and $\rho$ can be obtained by taking average over all configurations in Eq.\eqref{eq:operatorevol} and Eq.\eqref{eq:stateevol}. Thus,
	\begin{eqnarray}
		\dfrac{d}{dt}\opA(t)	&=&	i\left[\opHb,\opA(t) \right] + i\covdiff{t}{\opHb}{\opA}^\dagger;\label{eq:operatormeanevol}\\
		\dfrac{d}{dt}\rho(t)	&=&	-i\left[\opHb,\rho(t) \right] - i\covdiff{t}{\opHb}{\rho},\label{eq:statemeanevol}
	\end{eqnarray}
where $\opA(t)$ and $\rho(t)$ denote the $\omega$-average of $\opA_\omega(t)$ and $\rho_\omega(t)$, respectively; and 
	\begin{equation}
		\covdiff{t}{\opHb}{\cdot} := \Ex{\left[\delta\opHb_\omega(t),\mapUf(t)(\cdot) \right]}
		\label{eq:corr}
	\end{equation}
with $\delta\opHb_\omega(t):=\opHf(t)-\opHb$ and $\mapUf(t)$ being unitary evolution map from the initial time $t=0$ to time $t$ corresponding to Eq.\eqref{eq:stateevol} for a specific configuration $\omega$, is a non-homogeneous contribution from the mean-deviation of the configurations about the mean Hamiltonian $\opHb$. One can see that the argument in Eq. \eqref{eq:corr} is a Schr\"odinger operator and also an initial state of the evolution in the previous system of equations Eqs. \eqref{eq:operatorevol}--\eqref{eq:stateevol}. Moreover, from Eq.\eqref{eq:statemeanevol}, one can see that the term $\covdiff{t}{\opHb}{\rho} = \covdiff{t}{\opVb}{\rho}$ contains the information of dependence between the Hamiltonian and the density operator or the state. $\covdiff{t}{\opHb}{\cdot}$ vanishes when $\delta\opHb_\omega(t)$ commute with  $\opA_\omega(t)$ and $\rho_\omega(t)$ at any time $t$.

\section{Modified Kubo's Formula and Fluctuation and Dissipation Theorem}
\subsection{Effects from the Bath}
From Eq.\eqref{eq:statemeanevol}, one obtains an ensemble average of a physical quantity corresponding to the operator $\opA$ by employing the equivalence between the Heisenberg and Schr\"odinger pictures, so one need investigate only the dynamics of the state $\rho$. Here we will consider the linear response formulation as define in Ref.\cite{1966Kubo} to find the behaviour of the system when it is weakly contacted with a bath. First of all, we define a small interaction term with the bath as 
\begin{equation} \LI^t(\cdot) := \left[\opB, \cdot \right]\h(t),
\end{equation} 
where $\opB$ is an operator representing the effects from the bath, and $\h$ is its corresponding $c$-number. Adding this term to Eq.\eqref{eq:statemeanevol} leads to a perturbed evolution\footnote{For the detailed derivation, see the Appendix \ref{sec:detail}.} as
	\begin{equation}
		\dfrac{d}{dt}\rho(t)	=	-i\left(\Lb - \LI^t\right)\rho(t) - i\covdiff{t}{\opHb}{\rho},
		\label{eq:perturbed}
	\end{equation}
where $\Lb$ is the Liouville operator corresponding to the mean Hamiltonian $\opHb$. Its formal solution can be written as \cite{1998Ingold}
	\begin{eqnarray}
		\rho(t)	&=&	e^{-it\Lb}\rho(0) + i\int_0^te^{-i(t-s)\Lb}\LI^s\rho(s) ds \nonumber\\
				& & +\left[\eta(t) + i\int_0^te^{-i(t-s)\Lb}\LI^s\eta(s) ds  \right]\nonumber\\
				& & + \mathcal{O}(h_\opB^2),
		\label{eq:formalstate}
	\end{eqnarray}
where 
	\begin{equation}
		\eta(t) = i\int_0^te^{-i(t-s)\Lb}\covdiff{s}{\opHb}{\rho} ds.
	\end{equation}
In the case, the initial state $\rho(0)$ is an adjusted equilibrium state $\equi$, so it follows that $e^{-it\Lb}\equi = \equi$. Combining Eqs.\eqref{eq:formalstate} and \eqref{eq:Schro-Heiequivalence} and performing iteration to calculate the average of the observable $\opA$;
	\begin{eqnarray}
		\aveb{\opA(t)}_\rho &=& \avef{\opA} + i\int_0^t\left(  \tr\left(\equi\left[\opA(t),\opB(s)\right]\right)h_\opB(s) \right)ds\nonumber\\
		& &+\tr\left[\eta(t)\opA + i\int_0^te^{-i(t-s)\Lb}\LI^s\eta(s)\opA ds  \right]\nonumber\\ 
		& &+\mathcal{O}(h_\opB^2).
		\label{eq:Aaveragedynamic}
	\end{eqnarray}
From the expression above, one can see that an average of physical quantity is deviated not only by a perturbation from a bath solely (the second term on the right hand side), but also by the non-homogeneous term $\eta(t)$ in the third term and coupling between the bath and random effects in the forth term. In the sense that the bath is a measurement probe, a linear response which is related to a measured quantity will be deviated by those effects also.

\subsection{Modified Kubo's formula and Fluctuation Dissipation Relation}\label{sec:mFDR}
The linear response function is defined by
	
	\begin{equation}
	\resp{\opA}{\opB}(t,t') = \dfrac{\partial \aveb{\opA(t)}_\rho}{\partial \h(t')}\bigg\vert_{\h=0}.
	\label{eq:response}
	\end{equation}	
Then we obtain, for $0\leq t'\leq t$,
	\begin{equation}
		\hspace{-.5cm}\resp{\opA}{\opB}(t,t') = 2i\theta(t-t')\{\tr\left(\equi\left[\opA(t),\opB(t')\right]\right) + \Delta_{\opA\opB}(t,t') \},
	\end{equation}	
	where
	\begin{align}
	\Delta_{\opA\opB}(t,t') &:= \dfrac{1}{2}\tr\left[\left(\int_0^{t'} e^{-i(t-t')\Lb}\right.\right.\nonumber\\
		&\times \left.\left.\left[\opB,e^{-i(t'-s)\Lb} \covdiff{s}{\opHb}{\rho} \right] ds\right)\opA\right].
	\label{eq:extraterm}
	\end{align}
In physical sense, the term $\tr\left(\equi\left[\opA(t),\opB(t')\right]\right)$ can be related to an symmetric correlation \cite{2011Cugliandolo}, where correlation is usually defined as a two-epoch time correlation function between operators $\opA$ and $\opB$ in a given state. In this case:
	\begin{equation}
		\corr{\opA}{\opB}\left(t,t'\right) := \avef{\opA(t)\opB(t')}.
	\end{equation}
	One can then define symmetric and anti-symmetric correlations respectively by
	\begin{eqnarray}
		\corr{\opA}{\opB}^{-}\left(t,t'\right) &:=& \avef{ \left[\opA(t),\opB(t')\right]}, \\
		\corr{\opA}{\opB}^{+}\left(t,t'\right) &:=& \avef{ \{\opA(t),\opB(t')\} },
	\end{eqnarray}
where $\{\opA,\opB\}:= \opA\opB + \opB\opA$. Thus, the modified form of the Kubo's formula can be written as
	\begin{equation}
	\label{eq:Kubo}
		\resp{\opA}{\opB}(t,t') = 2i\theta(t-t')\left[\corr{\opA}{\opB}^{-}\left(t,t'\right) + \Delta_{\opA\opB}(t,t') \right].
	\end{equation}
Note that, from a typical set-up of linear response, $\h(t)$ is set to be zero for $t<t'$ and then turned on at $t'$ to reserve the causality of the measurement -- the effect from the probe cannot influence the system before they are in contact at the time $t=t'$ \cite{1998Ingold,2010speck}. In other words, from the equation it can be said that the system is in adjusted equilibrium state at the initial time $t = 0$ and evolves until $t = t'$. Then, it makes weakly contact with the bath (which can well be a measurement probe), and they evolve together until the present time $t$. All information of the evolution of the system and of the concerned operator is therefore contained in Eq.\eqref{eq:Kubo}.
	
Now consider the conditions to obtain the original FDR from Eq.\eqref{eq:Kubo}. Using the short-hand writing $(t,t')\rightarrow(t)$ when $t'=0$, so that $\corr{\opA}{\opB}\left(t,t'\right) \rightarrow \corr{\opA}{\opB}\left(t\right)$, 
we obtain
	\begin{eqnarray}
		\corr{\opA}{\opB}\left(t\right)	&=&	\tr\left(\equi\opA(t)\opB \right)\nonumber\\
				&=&	\dfrac{\tr\left(\opUb(t)\opB\opUb^\dagger(i\beta)\opUb^\dagger(t)\opA \right)}{\tr\opUb(i\beta)}\nonumber\\
		 \corr{\opA}{\opB}\left(t\right)	&=& \corr{\opB}{\opA}\left(-t-i\beta\right).
		\label{eq:commutincorr}
	\end{eqnarray}
One can see that the relation above is another version of the Kubo-Martin-Schwinger(KMS) condition, and the evolution along imaginary time arises here.

Suppose the interaction of the composite system with the bath begins at $t'= 0$ (or equivalently, the probe start at time $t = 0$ when the system is still in adjusted equilibrium state.) Then one can compute that $\Delta_{\opA\opB}(t) = 0$ \footnote{Here, one can see that the resulting equation can be approached by another reason or assumption, for instance, when one forces the system so that $\rho_\omega(t)$ always lies on its mean state $\rho(t)$ -- which is an alternative expression for \emph{mean-field approximation} also.}; and Eq.\eqref{eq:Kubo} becomes 
	\begin{equation}
	\resp{\opA}{\opB}(t) = 2i\theta(t)\corr{\opA}{\opB}^{-}(t).
	\label{eq:bFDR}
	\end{equation}
Taking the Fourier transform 
of the last equation, one obtains
	\begin{eqnarray}
		\hresp{\opA}{\opB}(\lambda) &=& \hcorr{\opA}{\opB}^{-}(\lambda) =	\left(1-e^{-\beta\lambda} \right)\hcorr{\opA}{\opB}(\lambda).
	\end{eqnarray}
The last relation is a direct consequence of Eq.\eqref{eq:commutincorr}. Furthermore, one finds that 
	\begin{equation}
		\hresp{\opA}{\opB}(\lambda) = \tanh\left(\dfrac{\beta\lambda}{2}\right)\hcorr{\opA}{\opB}^{+}(\lambda)
	\label{eq:FDR}
	\end{equation}
	which is similar to the \emph{standard} fluctuation-dissipation relation in the sense that it is calculated from a Gibbs state \cite{1966Kubo,1998Ingold}, but with a modification that the adjusted equilibrium state $\equi$ contains the information of randomness. Nonetheless, notice that only the average (wrt $\omega$) of the random potential effects to the adjusted equilibrium state, resulting the shift to the measured energy. For the case that $\opVb=0$, despite having external $\omega$-fluctuation, the adjusted equilibrium $\equi$ become coincide with $\gibbstate$ and the last relation is exactly the standard FDR.

\section{Discussion and Conclusion}
\paragraph{Connection to Other Works}
We revisit the model under consideration: $\opHf(\x,t) = \opHn + \opV(\x,t)$, where $\opHn$ is a constant or fixed Hamiltonian, and $\opV(\x,t)$ is multiplicative operator by a random smooth function $\fncV(\x,t)$. The presence of a random external random potential introduces unknown degrees of freedom. Unlike the bath or reservoir model, all dynamical properties of the external system $\sysV$ such as the relaxation time, dimension or volume are not specified -- so that the system can evolve to equilibrium as contrary to the conventional formulation of bath. This idea is analogous to having unobserved degrees of freedom. Similar model was analysed by Budini et al in Ref. \cite{2005Budini} to derive the master equation of non-Markovian type. It suggests that the presence of additional degrees of freedom, random disturbance to the Hamiltonian of the composite system, provides the memory effect in the master equation \cite{2005Budini}. In this work, instead of using the Hermitian operators to capture the unobserved degrees of freedom associated with the system with random coupling constant, the potential is employed in form of time-dependent random operator. One can formulate similar same scheme by setting that $\opV(\x,t) = \lambda_\omega\mathbf{Q}$, where $\mathbf{Q}$ is a Hermitian operator, and $\lambda_\omega$ is a random real number representing the strength of the coupling, and by using the non-Markovian master equation instead of Eq.\eqref{eq:statemeanevol}. It can be seen that the potential can be reduced to time-independent, and the adjusted equilibrium state can be defined because the a $\omega$-average of the Hamiltonian yields a constant of motion in this case. The extension in the direction of the present work is still open in general, especially for physical interpretations and implementations.

\paragraph{Connection to Stochastic Path Integral Formalism}
It is observed here that, as the commutator obeys the Lie structure, the term $\covdiff{t}{\opHb}{\rho}$ can be analysed as a derivative. In this instance, the deviated Hamiltonian $\delta\opHf(t):=\opHf(t)-\opHb$ appearing in $\covdiff{t}{\opHb}{\rho}$ will generate a dynamics for an appropriate dynamical parameter, and the derivative (concerning a commutation) represents a change of quantities related to randomness. For example, let consider a naive form, for a particular configuration $\omega$,
\[
	\left[\delta\opHf(t),\rho_\omega(t)-\rho(t) \right].
\]
Since a trace class $\mathcal{B}_1(\spacen)=\{\opA\in\mathcal{B}(\spacen):\tr(\opA)<\infty\}$, where $\mathcal{B}(\spacen)$ is an algebra of all bounded operator on $\spacen$, together with Hilbert-Schmidt inner product $(\opA,\opB)=\tr\left[\opA^\dagger\opB\right]$ forms a separable Hilbert space $\breve{\spacen}$ \cite[p.33]{mathlang}, clearly, the set of density operators is contained in $\mathcal{B}(\spacen)$ or $\breve{\spacen}$, and any density operator $\rho$ can be viewed as a vector in $\breve{\spacen}$ while an action $\mathcal{R}_\omega(\cdot):=i\left[\delta\opHf(t),\cdot \right]$ is an operator on $\breve{\spacen}$. Let $\mathcal{B}(\breve{\spacen})$ denote a Banach space of all bounded operators on $\breve{\spacen}$, one can find that the inner product 
\[
\left(\rho,\mathcal{R}_\omega(\rho)\right)=i\left(\tr\rho\left[\delta\opHf(t),\rho \right]\right)=0
\]
following the cyclic invariance of trace function. Thus, by Lumer-Phillips Theorem, the operator $\mathcal{R}_\omega(\cdot)$ is dissipative and can be a generator of contraction semigroup on $\breve{\spacen}$ \cite[Theorem 2.5]{Rivas}. However, the semigroup given here is determined up to the configuration $\omega$, so does a dynamical parameter (denoted by $\tau_\omega$) of the evolution. Therefore, we will obtain
\begin{equation}
	\dfrac{\partial\rho}{\partial\tau_\omega}=\mathcal{R}_\omega(\rho),
\end{equation}
another governing equation for density operator $\rho$. Hence, Eq.\eqref{eq:operatormeanevol} and Eq.\eqref{eq:statemeanevol}, for the case of countable number of configurations case, become multi-dimensional problem (in parameters). When the cardinality of $\Omega$ is uncountable, the description of the problem in this direction will be considered as a version of path-integral or stochastic dynamics. Indeed, derivations in those fields can be adopted to analyse the problem in this scheme, and still remain open. 

\paragraph{Remark on Equilibrium State} 
By the definition of the external potential alone, it can be seen that the dynamics of the system looks suspicious because it may not reach the equilibrium within a finite time. In fact, the equilibrium state, which maximizes the von Neumann entropy, is not well defined in this case because of lacking of information all over the time domain. Even though the Gibbs' state can be defined for specific Hamiltonian at specific time, the time-dependence of the Hamiltonian causes ambiguity to the model by the fact that the Gibbs' state defined here is dependent on time following the Hamiltonian. That says, when the Hamiltonian changes, the preferred direction of the dynamics in the state space also changes. Thus, the equilibrium condition depends on time and the dynamics as equilibration becomes questionable. In order to answer the question, one idea is that one can define an instantaneous equilibrium state which is given by the Hamiltonian at instant time and for specific configuration $\omega$. Then, the dynamics can be considered from a trajectory of such instantaneous equilibrium states, and the deviation from that state involves the entropy production along the path dynamics \cite{PhysRevLett.107.140404}. On the contrary, in this work, by employing the assumption \eqref{cond:constV} that the energy corresponding to the $\omega$-average of Hamiltonian is a constant of motion, and one can define a single adjusted equilibrium state, irregarding to time, as in this formulation. Hence, the mean dynamics of the state can be derived from the equilibrium condition.

\paragraph{Remark on Entropy Production}
There remains interesting topic involving the entropy and information in views of the entropy function $S(\rho) := -\tr\rho\ln\rho$. It is already known that the evolution of the system for a fixed configuration $\omega$ is unitary, but the unknown random potential still affects the dynamics. Thus, one cannot know exactly in which the state the system lies. When we reduce the dynamics to the mean dynamics as in Eq.\eqref{eq:statemeanevol}, we expect entropy change by the averaging evolution process. To demonstrate this issue, consider Eq.\eqref{eq:stateevol}, one can see that their entropies are unchanged along the time by invariance under the unitary transformation of the entropy function \cite{RevModPhys.50.221}. Because the entropy function is concave, it follows that
	\begin{align*}
		S(\rho(t)) &= S(\Ex{\rho_\omega(t)})\geq \Ex{S(\rho_\omega(t))}\\
		 &= \Ex{S(\rho(0))} = S(\rho(0)).
	\end{align*}
Then, the margin of the entropies $S(\rho(t))-S(\rho(0))$, which can be treated as the overall entropy production along the dynamics until the time $t$, can be positive. Additionally, for the case that $\covdiff{t}{\opHb}{\rho}=0$ which is occur, for instant, when the random effect is absent; one can define an entropy production \cite{JaksicPillet2001} as
	\begin{equation}
		Ent_\opB(t):=-i\beta
		h_\opB(t)\tr\rho(t)\left[\opHb,\opB\right],
		\label{eq:entropyproduction}
	\end{equation}
since the effect from the bath is treated as a perturbation and by choosing the adjusted equilibrium state $\equi$ as a reference state. Because the entropy production above is in linear order of $h_\opB$, the evolution is still close to the adiabatic regime. On the contrary, for the case that $\covdiff{t}{\opHb}{\rho}\neq0$, the evolution map corresponding to Eq.\eqref{eq:statemeanevol} may not admit a group or even a semi-group properties in general. The definition Eq.\eqref{eq:entropyproduction} is therefore not well defined for this case, and one can still expect a non-adiabatic effect in some situation.


\paragraph{Remark on the Extra Term}
According to the derivation of the modified FDR in Section \ref{sec:mFDR}, it is an advantage of using the mean dynamics Eq.\eqref{eq:statemeanevol}. As a result, it coincides with many previous works of the modified FDR: the correlation in the former FDR is replaced by a composition of the old one and another extra term. It was often suggested in literature that the function of the extra term is to explain the behaviour of the considered physical system out of the equilibrium regime. One idea is that the extra contribution to the response function is due to the dynamical activity, the time-symmetric part of action from the bath (or external system), which usually vanishes in equilibrium by the causality argument -- the response must not occur before a measurement \cite{PhysRevE.87.022125}. However, when the system is out of equilibrium or there are other parameters not included in the consideration, the dynamical activity seems to cause an amount of energy dissipation \cite{Baiesi2009,Baiesi2010,Baiesi2013,PhysRevLett.112.140602}.  It has been also interpreted as total entropy production from the dynamics, while the former term is treated as the entropy production of the medium or the considered system \cite{2010speck,Seifert2010}. In our case, the extra term arises from the presence of
\[\covdiff{t}{\opHb}{\rho} = \Ex{\left[\delta\opHf(t),\rho_\omega(t) \right]}\]
in the governing equation of the mean dynamics, where $\delta\opHf(t) := \opHf(t) - \opHb$. One can see that it keeps all the information of the external random potential at any time through which the system evolves. Recall Eq.\eqref{eq:extraterm} for the explicit form of the extra term in the Kubo's formula 
	\begin{align}
	\Delta_{\opA\opB}(t,t') &:= \dfrac{1}{2}\tr\left[\left(\int_0^{t'} e^{-i(t-t')\Lb}\right.\right.\nonumber\\
		&\times \left.\left.\left[\opB,e^{-i(t'-s)\Lb}\covdiff{s}{\opHb}{\rho} \right] ds\right)\opA\right].
	\end{align}
It can be seen that the actions along the dynamics are all included in the expression at the point they act. There are two evolutions therein: one maps from the initial time to the measurement time accumulating the effects from the non-homogeneous term $\covdiff{s}{\opHb}{\rho}$ interacted with a bath via the operator $\opB$ along the time, and the second one $e^{-i(t-t')\Lb}$ makes the system and bath evolve until the final time $t$. Therefore, it reflects the memory effect of the whole evolution, equipped with the non-homogeneous term from the random potential, in the measured quantity. In particular, the added random term disturbs the measured quantity to be deviate from a considered value in any experiment it involved with unless in the limit $t'\rightarrow 0$; making the state in the adjusted equilibrium state at the measurement time as shown in Eq.\eqref{eq:bFDR}. 

\paragraph{Conclusion}
In summary, we find that the deviation of the FDR can be designed as an effect from a randomness. By considering a class of Schr\"odinger type equations indexed by configurations in a probability space and taking an configuration-expectation, the mean dynamics with non-homogeneous term will be obtained. A perturbation from a bath in linear order yield a deviation of average of a physical quantity which affects a linear response. Consequently, the modified-FDR is derived and, we find that it also relax to the standard FDR in the case that the probe begins to contact with the system when the latter has not yet reached an equilibrium, but is in an adjusted equilibrium.

\begin{acknowledgments}
F. Sakuldee would like to thank Sri-Trang Thong Scholarship, Faculty of Science, Mahidol University for financial support to study at Department of Physics, Faculty of Science, Mahidol University.
\end{acknowledgments}

\bibliography{mybib}

\begin{thebibliography}{40}%
\makeatletter
\providecommand \@ifxundefined [1]{%
 \@ifx{#1\undefined}
}%
\providecommand \@ifnum [1]{%
 \ifnum #1\expandafter \@firstoftwo
 \else \expandafter \@secondoftwo
 \fi
}%
\providecommand \@ifx [1]{%
 \ifx #1\expandafter \@firstoftwo
 \else \expandafter \@secondoftwo
 \fi
}%
\providecommand \natexlab [1]{#1}%
\providecommand \enquote  [1]{``#1''}%
\providecommand \bibnamefont  [1]{#1}%
\providecommand \bibfnamefont [1]{#1}%
\providecommand \citenamefont [1]{#1}%
\providecommand \href@noop [0]{\@secondoftwo}%
\providecommand \href [0]{\begingroup \@sanitize@url \@href}%
\providecommand \@href[1]{\@@startlink{#1}\@@href}%
\providecommand \@@href[1]{\endgroup#1\@@endlink}%
\providecommand \@sanitize@url [0]{\catcode `\\12\catcode `\$12\catcode
  `\&12\catcode `\#12\catcode `\^12\catcode `\_12\catcode `\%12\relax}%
\providecommand \@@startlink[1]{}%
\providecommand \@@endlink[0]{}%
\providecommand \url  [0]{\begingroup\@sanitize@url \@url }%
\providecommand \@url [1]{\endgroup\@href {#1}{\urlprefix }}%
\providecommand \urlprefix  [0]{URL }%
\providecommand \Eprint [0]{\href }%
\providecommand \doibase [0]{http://dx.doi.org/}%
\providecommand \selectlanguage [0]{\@gobble}%
\providecommand \bibinfo  [0]{\@secondoftwo}%
\providecommand \bibfield  [0]{\@secondoftwo}%
\providecommand \translation [1]{[#1]}%
\providecommand \BibitemOpen [0]{}%
\providecommand \bibitemStop [0]{}%
\providecommand \bibitemNoStop [0]{.\EOS\space}%
\providecommand \EOS [0]{\spacefactor3000\relax}%
\providecommand \BibitemShut  [1]{\csname bibitem#1\endcsname}%
\let\auto@bib@innerbib\@empty
\bibitem [{\citenamefont {Callen}\ and\ \citenamefont
  {Welton}(1951)}]{1951CallenWelton}%
  \BibitemOpen
  \bibfield  {author} {\bibinfo {author} {\bibfnamefont {H.~B.}\ \bibnamefont
  {Callen}}\ and\ \bibinfo {author} {\bibfnamefont {T.~A.}\ \bibnamefont
  {Welton}},\ }\href@noop {} {\bibfield  {journal} {\bibinfo  {journal}
  {Physical Review}\ }\textbf {\bibinfo {volume} {83}},\ \bibinfo {pages} {34 }
  (\bibinfo {year} {1951})}\BibitemShut {NoStop}%
\bibitem [{\citenamefont {Kubo}(1966)}]{1966Kubo}%
  \BibitemOpen
  \bibfield  {author} {\bibinfo {author} {\bibfnamefont {R.}~\bibnamefont
  {Kubo}},\ }\href@noop {} {\bibfield  {journal} {\bibinfo  {journal} {Progress
  Report on Theoretical Physics}\ }\textbf {\bibinfo {volume} {29}},\ \bibinfo
  {pages} {255 } (\bibinfo {year} {1966})}\BibitemShut {NoStop}%
\bibitem [{\citenamefont {Crisanti}\ and\ \citenamefont
  {Ritort}(2003)}]{2003Crisanti}%
  \BibitemOpen
  \bibfield  {author} {\bibinfo {author} {\bibfnamefont {A.}~\bibnamefont
  {Crisanti}}\ and\ \bibinfo {author} {\bibfnamefont {F.}~\bibnamefont
  {Ritort}},\ }\href {http://stacks.iop.org/0305-4470/36/i=21/a=201} {\bibfield
   {journal} {\bibinfo  {journal} {Journal of Physics A: Mathematical and
  General}\ }\textbf {\bibinfo {volume} {36}},\ \bibinfo {pages} {R181}
  (\bibinfo {year} {2003})}\BibitemShut {NoStop}%
\bibitem [{\citenamefont {Cugliandolo}(2011)}]{2011Cugliandolo}%
  \BibitemOpen
  \bibfield  {author} {\bibinfo {author} {\bibfnamefont {L.~F.}\ \bibnamefont
  {Cugliandolo}},\ }\href {http://stacks.iop.org/1751-8121/44/i=48/a=483001}
  {\bibfield  {journal} {\bibinfo  {journal} {Journal of Physics A:
  Mathematical and Theoretical}\ }\textbf {\bibinfo {volume} {44}},\ \bibinfo
  {pages} {483001} (\bibinfo {year} {2011})}\BibitemShut {NoStop}%
\bibitem [{\citenamefont {Cugliandolo}\ \emph {et~al.}(1997)\citenamefont
  {Cugliandolo}, \citenamefont {Kurchan}, \citenamefont {{Le Doussal}},\ and\
  \citenamefont {Peliti}}]{PhysRevLett.78.350}%
  \BibitemOpen
  \bibfield  {author} {\bibinfo {author} {\bibfnamefont {L.~F.}\ \bibnamefont
  {Cugliandolo}}, \bibinfo {author} {\bibfnamefont {J.}~\bibnamefont
  {Kurchan}}, \bibinfo {author} {\bibfnamefont {P.}~\bibnamefont {{Le
  Doussal}}}, \ and\ \bibinfo {author} {\bibfnamefont {L.}~\bibnamefont
  {Peliti}},\ }\href {\doibase 10.1103/PhysRevLett.78.350} {\bibfield
  {journal} {\bibinfo  {journal} {Phys. Rev. Lett.}\ }\textbf {\bibinfo
  {volume} {78}},\ \bibinfo {pages} {350} (\bibinfo {year} {1997})}\BibitemShut
  {NoStop}%
\bibitem [{\citenamefont {Grigera}\ and\ \citenamefont
  {Israeloff}(1999)}]{PhysRevLett.83.5038}%
  \BibitemOpen
  \bibfield  {author} {\bibinfo {author} {\bibfnamefont {T.~S.}\ \bibnamefont
  {Grigera}}\ and\ \bibinfo {author} {\bibfnamefont {N.~E.}\ \bibnamefont
  {Israeloff}},\ }\href {\doibase 10.1103/PhysRevLett.83.5038} {\bibfield
  {journal} {\bibinfo  {journal} {Phys. Rev. Lett.}\ }\textbf {\bibinfo
  {volume} {83}},\ \bibinfo {pages} {5038} (\bibinfo {year}
  {1999})}\BibitemShut {NoStop}%
\bibitem [{\citenamefont {Ricci-Tersenghi}\ \emph {et~al.}(2000)\citenamefont
  {Ricci-Tersenghi}, \citenamefont {Stariolo},\ and\ \citenamefont
  {Arenzon}}]{PhysRevLett.84.4473}%
  \BibitemOpen
  \bibfield  {author} {\bibinfo {author} {\bibfnamefont {F.}~\bibnamefont
  {Ricci-Tersenghi}}, \bibinfo {author} {\bibfnamefont {D.~A.}\ \bibnamefont
  {Stariolo}}, \ and\ \bibinfo {author} {\bibfnamefont {J.~J.}\ \bibnamefont
  {Arenzon}},\ }\href@noop {} {\bibfield  {journal} {\bibinfo  {journal} {Phys.
  Rev. Lett.}\ }\textbf {\bibinfo {volume} {84}},\ \bibinfo {pages} {4473}
  (\bibinfo {year} {2000})}\BibitemShut {NoStop}%
\bibitem [{\citenamefont {Cugliandolo}\ and\ \citenamefont
  {Kurchan}(1993)}]{PhysRevLett.71.173}%
  \BibitemOpen
  \bibfield  {author} {\bibinfo {author} {\bibfnamefont {L.~F.}\ \bibnamefont
  {Cugliandolo}}\ and\ \bibinfo {author} {\bibfnamefont {J.}~\bibnamefont
  {Kurchan}},\ }\href {\doibase 10.1103/PhysRevLett.71.173} {\bibfield
  {journal} {\bibinfo  {journal} {Phys. Rev. Lett.}\ }\textbf {\bibinfo
  {volume} {71}},\ \bibinfo {pages} {173} (\bibinfo {year} {1993})}\BibitemShut
  {NoStop}%
\bibitem [{\citenamefont {Bellon}\ \emph {et~al.}(2001)\citenamefont {Bellon},
  \citenamefont {Ciliberto},\ and\ \citenamefont {Laroche}}]{Bellon2001}%
  \BibitemOpen
  \bibfield  {author} {\bibinfo {author} {\bibfnamefont {L.}~\bibnamefont
  {Bellon}}, \bibinfo {author} {\bibfnamefont {S.}~\bibnamefont {Ciliberto}}, \
  and\ \bibinfo {author} {\bibfnamefont {C.}~\bibnamefont {Laroche}},\ }\href
  {http://stacks.iop.org/0295-5075/53/i=4/a=511} {\bibfield  {journal}
  {\bibinfo  {journal} {EPL (Europhysics Letters)}\ }\textbf {\bibinfo {volume}
  {53}},\ \bibinfo {pages} {511} (\bibinfo {year} {2001})}\BibitemShut
  {NoStop}%
\bibitem [{\citenamefont {Cugliandolo}\ \emph {et~al.}(1999)\citenamefont
  {Cugliandolo}, \citenamefont {Grempel}, \citenamefont {Kurchan},\ and\
  \citenamefont {Vincent}}]{Cuglian1999}%
  \BibitemOpen
  \bibfield  {author} {\bibinfo {author} {\bibfnamefont {L.~F.}\ \bibnamefont
  {Cugliandolo}}, \bibinfo {author} {\bibfnamefont {D.~R.}\ \bibnamefont
  {Grempel}}, \bibinfo {author} {\bibfnamefont {J.}~\bibnamefont {Kurchan}}, \
  and\ \bibinfo {author} {\bibfnamefont {E.}~\bibnamefont {Vincent}},\ }\href
  {http://stacks.iop.org/0295-5075/48/i=6/a=699} {\bibfield  {journal}
  {\bibinfo  {journal} {EPL (Europhysics Letters)}\ }\textbf {\bibinfo {volume}
  {48}},\ \bibinfo {pages} {699} (\bibinfo {year} {1999})}\BibitemShut
  {NoStop}%
\bibitem [{\citenamefont {Marinari}\ \emph {et~al.}(1998)\citenamefont
  {Marinari}, \citenamefont {Parisi}, \citenamefont {Ricci-Tersenghi},\ and\
  \citenamefont {Ruiz-Lorenzo}}]{1998Enzo}%
  \BibitemOpen
  \bibfield  {author} {\bibinfo {author} {\bibfnamefont {E.}~\bibnamefont
  {Marinari}}, \bibinfo {author} {\bibfnamefont {G.}~\bibnamefont {Parisi}},
  \bibinfo {author} {\bibfnamefont {F.}~\bibnamefont {Ricci-Tersenghi}}, \ and\
  \bibinfo {author} {\bibfnamefont {J.~J.}\ \bibnamefont {Ruiz-Lorenzo}},\
  }\href {http://stacks.iop.org/0305-4470/31/i=11/a=011} {\bibfield  {journal}
  {\bibinfo  {journal} {Journal of Physics A: Mathematical and General}\
  }\textbf {\bibinfo {volume} {31}},\ \bibinfo {pages} {2611} (\bibinfo {year}
  {1998})}\BibitemShut {NoStop}%
\bibitem [{\citenamefont {Komatsu}\ \emph {et~al.}(2011)\citenamefont
  {Komatsu}, \citenamefont {L'H{\^o}te}, \citenamefont {Nakamae}, \citenamefont
  {Mosser}, \citenamefont {Konczykowski}, \citenamefont {Dubois}, \citenamefont
  {Dupuis},\ and\ \citenamefont {Perzynski}}]{PhysRevLett.106.150603}%
  \BibitemOpen
  \bibfield  {author} {\bibinfo {author} {\bibfnamefont {K.}~\bibnamefont
  {Komatsu}}, \bibinfo {author} {\bibfnamefont {D.}~\bibnamefont {L'H{\^o}te}},
  \bibinfo {author} {\bibfnamefont {S.}~\bibnamefont {Nakamae}}, \bibinfo
  {author} {\bibfnamefont {V.}~\bibnamefont {Mosser}}, \bibinfo {author}
  {\bibfnamefont {M.}~\bibnamefont {Konczykowski}}, \bibinfo {author}
  {\bibfnamefont {E.}~\bibnamefont {Dubois}}, \bibinfo {author} {\bibfnamefont
  {V.}~\bibnamefont {Dupuis}}, \ and\ \bibinfo {author} {\bibfnamefont
  {R.}~\bibnamefont {Perzynski}},\ }\href {\doibase
  10.1103/PhysRevLett.106.150603} {\bibfield  {journal} {\bibinfo  {journal}
  {Phys. Rev. Lett.}\ }\textbf {\bibinfo {volume} {106}},\ \bibinfo {pages}
  {150603} (\bibinfo {year} {2011})}\BibitemShut {NoStop}%
\bibitem [{\citenamefont {Rom{\'a}}\ \emph {et~al.}(2007)\citenamefont
  {Rom{\'a}}, \citenamefont {Bustingorry}, \citenamefont {Gleiser},\ and\
  \citenamefont {Dom{\'i}nguez}}]{PhysRevLett.98.097203}%
  \BibitemOpen
  \bibfield  {author} {\bibinfo {author} {\bibfnamefont {F.}~\bibnamefont
  {Rom{\'a}}}, \bibinfo {author} {\bibfnamefont {S.}~\bibnamefont
  {Bustingorry}}, \bibinfo {author} {\bibfnamefont {P.~M.}\ \bibnamefont
  {Gleiser}}, \ and\ \bibinfo {author} {\bibfnamefont {D.}~\bibnamefont
  {Dom{\'i}nguez}},\ }\href {\doibase 10.1103/PhysRevLett.98.097203} {\bibfield
   {journal} {\bibinfo  {journal} {Phys. Rev. Lett.}\ }\textbf {\bibinfo
  {volume} {98}},\ \bibinfo {pages} {097203} (\bibinfo {year}
  {2007})}\BibitemShut {NoStop}%
\bibitem [{\citenamefont {Stariolo}\ and\ \citenamefont
  {Cannas}(1999)}]{PhysRevB.60.3013}%
  \BibitemOpen
  \bibfield  {author} {\bibinfo {author} {\bibfnamefont {D.~A.}\ \bibnamefont
  {Stariolo}}\ and\ \bibinfo {author} {\bibfnamefont {S.~A.}\ \bibnamefont
  {Cannas}},\ }\href {\doibase 10.1103/PhysRevB.60.3013} {\bibfield  {journal}
  {\bibinfo  {journal} {Phys. Rev. B}\ }\textbf {\bibinfo {volume} {60}},\
  \bibinfo {pages} {3013} (\bibinfo {year} {1999})}\BibitemShut {NoStop}%
\bibitem [{\citenamefont {Mitra}\ and\ \citenamefont
  {Millis}(2005)}]{PhysRevB.72.121102}%
  \BibitemOpen
  \bibfield  {author} {\bibinfo {author} {\bibfnamefont {A.}~\bibnamefont
  {Mitra}}\ and\ \bibinfo {author} {\bibfnamefont {A.~J.}\ \bibnamefont
  {Millis}},\ }\href {\doibase 10.1103/PhysRevB.72.121102} {\bibfield
  {journal} {\bibinfo  {journal} {Phys. Rev. B}\ }\textbf {\bibinfo {volume}
  {72}},\ \bibinfo {pages} {121102} (\bibinfo {year} {2005})}\BibitemShut
  {NoStop}%
\bibitem [{\citenamefont {Verley}\ \emph {et~al.}(2011)\citenamefont {Verley},
  \citenamefont {Ch{\'e}trite},\ and\ \citenamefont {Lacoste}}]{Verley2011}%
  \BibitemOpen
  \bibfield  {author} {\bibinfo {author} {\bibfnamefont {G.}~\bibnamefont
  {Verley}}, \bibinfo {author} {\bibfnamefont {R.}~\bibnamefont
  {Ch{\'e}trite}}, \ and\ \bibinfo {author} {\bibfnamefont {D.}~\bibnamefont
  {Lacoste}},\ }\href {http://stacks.iop.org/1742-5468/2011/i=10/a=P10025}
  {\bibfield  {journal} {\bibinfo  {journal} {Journal of Statistical Mechanics:
  Theory and Experiment}\ }\textbf {\bibinfo {volume} {2011}},\ \bibinfo
  {pages} {P10025} (\bibinfo {year} {2011})}\BibitemShut {NoStop}%
\bibitem [{\citenamefont {Zamponi}\ \emph {et~al.}(2005)\citenamefont
  {Zamponi}, \citenamefont {Ruocco},\ and\ \citenamefont
  {Angelani}}]{PhysRevE.71.020101}%
  \BibitemOpen
  \bibfield  {author} {\bibinfo {author} {\bibfnamefont {F.}~\bibnamefont
  {Zamponi}}, \bibinfo {author} {\bibfnamefont {G.}~\bibnamefont {Ruocco}}, \
  and\ \bibinfo {author} {\bibfnamefont {L.}~\bibnamefont {Angelani}},\ }\href
  {\doibase 10.1103/PhysRevE.71.020101} {\bibfield  {journal} {\bibinfo
  {journal} {Phys. Rev. E}\ }\textbf {\bibinfo {volume} {71}},\ \bibinfo
  {pages} {020101} (\bibinfo {year} {2005})}\BibitemShut {NoStop}%
\bibitem [{\citenamefont {Averin}\ and\ \citenamefont
  {Pekola}(2010)}]{PhysRevLett.104.220601}%
  \BibitemOpen
  \bibfield  {author} {\bibinfo {author} {\bibfnamefont {D.~V.}\ \bibnamefont
  {Averin}}\ and\ \bibinfo {author} {\bibfnamefont {J.~P.}\ \bibnamefont
  {Pekola}},\ }\href {\doibase 10.1103/PhysRevLett.104.220601} {\bibfield
  {journal} {\bibinfo  {journal} {Phys. Rev. Lett.}\ }\textbf {\bibinfo
  {volume} {104}},\ \bibinfo {pages} {220601} (\bibinfo {year}
  {2010})}\BibitemShut {NoStop}%
\bibitem [{\citenamefont {Caso}\ \emph {et~al.}(2010)\citenamefont {Caso},
  \citenamefont {Arrachea},\ and\ \citenamefont {Lozano}}]{PhysRevB.81.041301}%
  \BibitemOpen
  \bibfield  {author} {\bibinfo {author} {\bibfnamefont {A.}~\bibnamefont
  {Caso}}, \bibinfo {author} {\bibfnamefont {L.}~\bibnamefont {Arrachea}}, \
  and\ \bibinfo {author} {\bibfnamefont {G.~S.}\ \bibnamefont {Lozano}},\
  }\href {\doibase 10.1103/PhysRevB.81.041301} {\bibfield  {journal} {\bibinfo
  {journal} {Phys. Rev. B}\ }\textbf {\bibinfo {volume} {81}},\ \bibinfo
  {pages} {041301} (\bibinfo {year} {2010})}\BibitemShut {NoStop}%
\bibitem [{\citenamefont {Chen}(2008)}]{chen2008}%
  \BibitemOpen
  \bibfield  {author} {\bibinfo {author} {\bibfnamefont {L.~Y.}\ \bibnamefont
  {Chen}},\ }\href {\doibase 10.1063/1.2992153} {\bibfield  {journal} {\bibinfo
   {journal} {The Journal of Chemical Physics}\ }\textbf {\bibinfo {volume}
  {129}},\ \bibinfo {eid} {144113} (\bibinfo {year} {2008})}\BibitemShut
  {NoStop}%
\bibitem [{\citenamefont {Mauri}\ and\ \citenamefont
  {Leporini}(2006)}]{mauri2006}%
  \BibitemOpen
  \bibfield  {author} {\bibinfo {author} {\bibfnamefont {R.}~\bibnamefont
  {Mauri}}\ and\ \bibinfo {author} {\bibfnamefont {D.}~\bibnamefont
  {Leporini}},\ }\href {http://stacks.iop.org/0295-5075/76/i=6/a=1022}
  {\bibfield  {journal} {\bibinfo  {journal} {EPL (Europhysics Letters)}\
  }\textbf {\bibinfo {volume} {76}},\ \bibinfo {pages} {1022} (\bibinfo {year}
  {2006})}\BibitemShut {NoStop}%
\bibitem [{\citenamefont {Lobaskin}\ and\ \citenamefont
  {Kehrein}(2006)}]{Lobaskin2006}%
  \BibitemOpen
  \bibfield  {author} {\bibinfo {author} {\bibfnamefont {D.}~\bibnamefont
  {Lobaskin}}\ and\ \bibinfo {author} {\bibfnamefont {S.}~\bibnamefont
  {Kehrein}},\ }\href {\doibase 10.1007/s10955-006-9055-5} {\bibfield
  {journal} {\bibinfo  {journal} {Journal of Statistical Physics}\ }\textbf
  {\bibinfo {volume} {123}},\ \bibinfo {pages} {301} (\bibinfo {year}
  {2006})}\BibitemShut {NoStop}%
\bibitem [{\citenamefont {Baiesi}\ \emph {et~al.}(2009)\citenamefont {Baiesi},
  \citenamefont {Maes},\ and\ \citenamefont {Wynants}}]{Baiesi2009}%
  \BibitemOpen
  \bibfield  {author} {\bibinfo {author} {\bibfnamefont {M.}~\bibnamefont
  {Baiesi}}, \bibinfo {author} {\bibfnamefont {C.}~\bibnamefont {Maes}}, \ and\
  \bibinfo {author} {\bibfnamefont {B.}~\bibnamefont {Wynants}},\ }\href
  {\doibase 10.1007/s10955-009-9852-8} {\bibfield  {journal} {\bibinfo
  {journal} {Journal of Statistical Physics}\ }\textbf {\bibinfo {volume}
  {137}},\ \bibinfo {pages} {1094} (\bibinfo {year} {2009})}\BibitemShut
  {NoStop}%
\bibitem [{\citenamefont {Baiesi}\ \emph {et~al.}(2010)\citenamefont {Baiesi},
  \citenamefont {Boksenbojm}, \citenamefont {Maes},\ and\ \citenamefont
  {Wynants}}]{Baiesi2010}%
  \BibitemOpen
  \bibfield  {author} {\bibinfo {author} {\bibfnamefont {M.}~\bibnamefont
  {Baiesi}}, \bibinfo {author} {\bibfnamefont {E.}~\bibnamefont {Boksenbojm}},
  \bibinfo {author} {\bibfnamefont {C.}~\bibnamefont {Maes}}, \ and\ \bibinfo
  {author} {\bibfnamefont {B.}~\bibnamefont {Wynants}},\ }\href {\doibase
  10.1007/s10955-010-9951-6} {\bibfield  {journal} {\bibinfo  {journal}
  {Journal of Statistical Physics}\ }\textbf {\bibinfo {volume} {139}},\
  \bibinfo {pages} {492} (\bibinfo {year} {2010})}\BibitemShut {NoStop}%
\bibitem [{\citenamefont {Baiesi}\ and\ \citenamefont
  {Maes}(2013)}]{Baiesi2013}%
  \BibitemOpen
  \bibfield  {author} {\bibinfo {author} {\bibfnamefont {M.}~\bibnamefont
  {Baiesi}}\ and\ \bibinfo {author} {\bibfnamefont {C.}~\bibnamefont {Maes}},\
  }\href {http://stacks.iop.org/1367-2630/15/i=1/a=013004} {\bibfield
  {journal} {\bibinfo  {journal} {New Journal of Physics}\ }\textbf {\bibinfo
  {volume} {15}},\ \bibinfo {pages} {013004} (\bibinfo {year}
  {2013})}\BibitemShut {NoStop}%
\bibitem [{\citenamefont {Seifert}\ and\ \citenamefont
  {Speck}(2010)}]{Seifert2010}%
  \BibitemOpen
  \bibfield  {author} {\bibinfo {author} {\bibfnamefont {U.}~\bibnamefont
  {Seifert}}\ and\ \bibinfo {author} {\bibfnamefont {T.}~\bibnamefont
  {Speck}},\ }\href {http://stacks.iop.org/0295-5075/89/i=1/a=10007} {\bibfield
   {journal} {\bibinfo  {journal} {EPL (Europhysics Letters)}\ }\textbf
  {\bibinfo {volume} {89}},\ \bibinfo {pages} {10007} (\bibinfo {year}
  {2010})}\BibitemShut {NoStop}%
\bibitem [{\citenamefont {Katz}(1967)}]{1967Katz}%
  \BibitemOpen
  \bibfield  {author} {\bibinfo {author} {\bibfnamefont {A.}~\bibnamefont
  {Katz}},\ }\href@noop {} {\emph {\bibinfo {title} {{Principles of Statistical
  Mechanics}}}}\ (\bibinfo  {publisher} {W. H. Freeman and Company},\ \bibinfo
  {year} {1967})\BibitemShut {NoStop}%
\bibitem [{\citenamefont {Wehrl}(1978)}]{RevModPhys.50.221}%
  \BibitemOpen
  \bibfield  {author} {\bibinfo {author} {\bibfnamefont {A.}~\bibnamefont
  {Wehrl}},\ }\href {\doibase 10.1103/RevModPhys.50.221} {\bibfield  {journal}
  {\bibinfo  {journal} {Rev. Mod. Phys.}\ }\textbf {\bibinfo {volume} {50}},\
  \bibinfo {pages} {221} (\bibinfo {year} {1978})}\BibitemShut {NoStop}%
\bibitem [{Note1()}]{Note1}%
  \BibitemOpen
  \bibinfo {note} {For the detailed derivation, see the Appendix \ref
  {sec:detail}.}\BibitemShut {Stop}%
\bibitem [{\citenamefont {G-L}(1998)}]{1998Ingold}%
  \BibitemOpen
  \bibfield  {author} {\bibinfo {author} {\bibfnamefont {I.}~\bibnamefont
  {G-L}},\ }\enquote {\bibinfo {title} {{Dissipative Quantum Systems}},}\ in\
  \href@noop {} {\emph {\bibinfo {booktitle} {{Quantum Transport and
  Dissipation}}}}\ (\bibinfo  {publisher} {Wiley-VCH},\ \bibinfo {year}
  {1998})\ Chap.~\bibinfo {chapter} {4}\BibitemShut {NoStop}%
\bibitem [{\citenamefont {Speck}(2010)}]{2010speck}%
  \BibitemOpen
  \bibfield  {author} {\bibinfo {author} {\bibfnamefont {T.}~\bibnamefont
  {Speck}},\ }\href {\doibase 10.1143/PTPS.184.248} {\bibfield  {journal}
  {\bibinfo  {journal} {Progress of Theoretical Physics Supplement}\ }\textbf
  {\bibinfo {volume} {184}},\ \bibinfo {pages} {248} (\bibinfo {year}
  {2010})},\ \Eprint
  {http://arxiv.org/abs/http://ptps.oxfordjournals.org/content/184/248.full.pdf+html}
  {http://ptps.oxfordjournals.org/content/184/248.full.pdf+html} \BibitemShut
  {NoStop}%
\bibitem [{Note2()}]{Note2}%
  \BibitemOpen
  \bibinfo {note} {Here, one can see that the resulting equation can be
  approached by another reason or assumption, for instance, when one forces the
  system so that $\rho _\omega (t)$ always lies on its mean state $\rho (t)$ --
  which is an alternative expression for \protect \emph {mean-field
  approximation} also.}\BibitemShut {Stop}%
\bibitem [{\citenamefont {Budini}\ and\ \citenamefont
  {Schomerus}(2005)}]{2005Budini}%
  \BibitemOpen
  \bibfield  {author} {\bibinfo {author} {\bibfnamefont {A.~A.}\ \bibnamefont
  {Budini}}\ and\ \bibinfo {author} {\bibfnamefont {H.}~\bibnamefont
  {Schomerus}},\ }\href {http://stacks.iop.org/0305-4470/38/i=42/a=006}
  {\bibfield  {journal} {\bibinfo  {journal} {Journal of Physics A:
  Mathematical and General}\ }\textbf {\bibinfo {volume} {38}},\ \bibinfo
  {pages} {9251} (\bibinfo {year} {2005})}\BibitemShut {NoStop}%
\bibitem [{\citenamefont {Heinosaari}\ and\ \citenamefont
  {Ziman}(2011)}]{mathlang}%
  \BibitemOpen
  \bibfield  {author} {\bibinfo {author} {\bibfnamefont {T.}~\bibnamefont
  {Heinosaari}}\ and\ \bibinfo {author} {\bibfnamefont {M.}~\bibnamefont
  {Ziman}},\ }\href@noop {} {\emph {\bibinfo {title} {{The Mathematical
  Language of Quantum Theory: From Uncertainty to Entanglement}}}}\ (\bibinfo
  {publisher} {Cambridge University Press},\ \bibinfo {year}
  {2011})\BibitemShut {NoStop}%
\bibitem [{\citenamefont {Rivas}\ and\ \citenamefont {Huelga}(2012)}]{Rivas}%
  \BibitemOpen
  \bibfield  {author} {\bibinfo {author} {\bibfnamefont {Â.}~\bibnamefont
  {Rivas}}\ and\ \bibinfo {author} {\bibfnamefont {S.~F.}\ \bibnamefont
  {Huelga}},\ }\href@noop {} {\emph {\bibinfo {title} {{Open Quantum Systems:An
  Introduction}}}},\ {SpringerBriefs in Physics}\ (\bibinfo  {publisher}
  {Springer Berlin Heidelberg},\ \bibinfo {year} {2012})\BibitemShut {NoStop}%
\bibitem [{\citenamefont {Deffner}\ and\ \citenamefont
  {Lutz}(2011)}]{PhysRevLett.107.140404}%
  \BibitemOpen
  \bibfield  {author} {\bibinfo {author} {\bibfnamefont {S.}~\bibnamefont
  {Deffner}}\ and\ \bibinfo {author} {\bibfnamefont {E.}~\bibnamefont {Lutz}},\
  }\href {\doibase 10.1103/PhysRevLett.107.140404} {\bibfield  {journal}
  {\bibinfo  {journal} {Phys. Rev. Lett.}\ }\textbf {\bibinfo {volume} {107}},\
  \bibinfo {pages} {140404} (\bibinfo {year} {2011})}\BibitemShut {NoStop}%
\bibitem [{\citenamefont {Jak\v{s}i{\'c}}\ and\ \citenamefont
  {Pillet}(2001)}]{JaksicPillet2001}%
  \BibitemOpen
  \bibfield  {author} {\bibinfo {author} {\bibfnamefont {V.}~\bibnamefont
  {Jak\v{s}i{\'c}}}\ and\ \bibinfo {author} {\bibfnamefont {C.-A.}\
  \bibnamefont {Pillet}},\ }\href@noop {} {\bibfield  {journal} {\bibinfo
  {journal} {Communications in Mathematical Physics}\ }\textbf {\bibinfo
  {volume} {217}},\ \bibinfo {pages} {285} (\bibinfo {year}
  {2001})}\BibitemShut {NoStop}%
\bibitem [{\citenamefont {Maes}\ \emph {et~al.}(2013)\citenamefont {Maes},
  \citenamefont {Safaverdi}, \citenamefont {Visco},\ and\ \citenamefont {van
  Wijland}}]{PhysRevE.87.022125}%
  \BibitemOpen
  \bibfield  {author} {\bibinfo {author} {\bibfnamefont {C.}~\bibnamefont
  {Maes}}, \bibinfo {author} {\bibfnamefont {S.}~\bibnamefont {Safaverdi}},
  \bibinfo {author} {\bibfnamefont {P.}~\bibnamefont {Visco}}, \ and\ \bibinfo
  {author} {\bibfnamefont {F.}~\bibnamefont {van Wijland}},\ }\href {\doibase
  10.1103/PhysRevE.87.022125} {\bibfield  {journal} {\bibinfo  {journal} {Phys.
  Rev. E}\ }\textbf {\bibinfo {volume} {87}},\ \bibinfo {pages} {022125}
  (\bibinfo {year} {2013})}\BibitemShut {NoStop}%
\bibitem [{\citenamefont {Lippiello}\ \emph {et~al.}(2014)\citenamefont
  {Lippiello}, \citenamefont {Baiesi},\ and\ \citenamefont
  {Sarracino}}]{PhysRevLett.112.140602}%
  \BibitemOpen
  \bibfield  {author} {\bibinfo {author} {\bibfnamefont {E.}~\bibnamefont
  {Lippiello}}, \bibinfo {author} {\bibfnamefont {M.}~\bibnamefont {Baiesi}}, \
  and\ \bibinfo {author} {\bibfnamefont {A.}~\bibnamefont {Sarracino}},\ }\href
  {\doibase 10.1103/PhysRevLett.112.140602} {\bibfield  {journal} {\bibinfo
  {journal} {Phys. Rev. Lett.}\ }\textbf {\bibinfo {volume} {112}},\ \bibinfo
  {pages} {140602} (\bibinfo {year} {2014})}\BibitemShut {NoStop}%
\bibitem [{\citenamefont {Sakurai}(1993)}]{sakurai1993}%
  \BibitemOpen
  \bibfield  {author} {\bibinfo {author} {\bibfnamefont {J.~J.}\ \bibnamefont
  {Sakurai}},\ }\href
  {http://www.amazon.com/exec/obidos/redirect?tag=citeulike07-20&path=ASIN/0201539292}
  {\emph {\bibinfo {title} {{Modern Quantum Mechanics (Revised Edition)}}}},\
  \bibinfo {edition} {1st}\ ed.\ (\bibinfo  {publisher} {Addison Wesley},\
  \bibinfo {year} {1993})\BibitemShut {NoStop}%
\end{thebibliography}%

\appendix
\section{Detailed Derivation}
\label{sec:detail}
For a fixed configuration $\omega$, the dynamics of the system in this specific realization $\omega$ is well defined formally since $\fncV(\x,t)$ is assumed to be a smooth function. Thus, the evolution can take the form 
\[\opUf(t,t') = \T\exp\left(-i\int_{t'}^t \opHf(\tau) d\tau\right)\]
where $\T$ is a time-ordering operator with its reverse time-ordering operator denoted by $\overline{\T}$ \cite{sakurai1993}; here, we set $\opUf(t,0):=\opUf(t)$ for simplicity. The evolution of the operator $\opA$ follows
	\begin{eqnarray}
		\dfrac{d}{dt}\opA_\omega(t)	&=&	i\Lf^t(\opA_\omega(t)) := i\left[\opHf,\opA_\omega(t) \right],\label{eq:app:operatorevol}\\
		\opA_\omega(t)	&=&	e^{i\T\int_0^t ds{\mathcal{L}_\omega^s}}\opA = \opUf^\dagger(t)\opA\opUf(t),
	\end{eqnarray}
with its dual dynamics
	\begin{eqnarray}
		\dfrac{d}{dt}\rho_\omega(t)	&=&	-i\Lf^t(\rho_\omega(t)) := -i\left[\opHf,\rho_\omega(t) \right],\label{eq:app:stateevol}\\
		\rho_\omega(t)	&=&	e^{-i\T\int_0^t ds{\mathcal{L}_\omega^s}}\rho = \opUf(t)\rho\opUf^\dagger(t).\label{eq:app:stateevol2}
	\end{eqnarray}
However, we investigate its mean dynamics rather than the class of equations above, where the mean evolutions in Eqs. \eqref{eq:app:operatorevol} and \eqref{eq:app:stateevol} are given by
\begin{eqnarray}
		\dfrac{d}{dt}\opA(t)	&=&	i\left[\opHb,\opA(t) \right] + i\covdiff{t}{\opHb}{\opA}^\dagger,\label{eq:app:operatormeanevol}\\
		\dfrac{d}{dt}\rho(t)	&=&	-i\left[\opHb,\rho(t) \right] - i\covdiff{t}{\opHb}{\rho},\label{eq:app:statemeanevol}
	\end{eqnarray}
where $\opA(t)$ and $\rho(t)$ denote the $\omega$-average of $\opA_\omega(t)$ and $\rho_\omega(t)$, respectively; and 
	\begin{equation}
		\covdiff{t}{\opHb}{\cdot} := \Ex{\left[\delta\opHb_\omega(t),\mapUf(t)(\cdot) \right]}
	\end{equation}
with $\delta\opHb_\omega(t):=\opHf(t)-\opHb$ and $\mapUf(t)(\cdot)=\opUf(t)\cdot\opUf^\dagger(t)$ being a unitary evolution map from the initial time $t=0$ to time $t$ corresponding to Eq.\eqref{eq:app:stateevol} for a specific configuration $\omega$. Indeed, $\covdiff{t}{\opHb}{\cdot}$ can be treated as a non-homogeneous contribution from the mean-deviation of the configurations about the mean Hamiltonian $\opHb$. For the Eq.\eqref{eq:app:statemeanevol} which we will mainly focus has a formal solution as 
	\begin{equation}
		\rho(t)	= e^{-it\Lb}\rho(0) + \eta_\rho(t),
	\end{equation}
where 
	\begin{equation}
		\eta_\rho(t) := i\int_0^te^{-i(t-s)\Lb}\covdiff{s}{\opHb}{\rho} ds.
	\end{equation}
One can see that, from the integral form above, the evolution map, in general, may not admit the semi-group properties by the presence of the non-homogeneous term $\covdiff{s}{\opHb}{\rho}$.

Next, consider a perturbation from a bath $\LI^t\cdot=h_\opB(t)\left[\opB,\cdot \right]$. The perturbed evolution can be written as
	\begin{equation}
		\dfrac{d}{dt}\rho(t)	=	-i\left(\Lb - \LI^t\right)\rho(t) - i\covdiff{t}{\opHb}{\rho},
		\label{eq:app:perturbed}
	\end{equation}
where $\Lb$ is the Liouville operator corresponding to the mean Hamiltonian $\opHb$. Its formal solution can be written as \cite{1998Ingold}
	\begin{eqnarray}
		\rho(t)	&=&	e^{-it\Lb}\rho(0) + i\int_0^te^{-i(t-s)\Lb}\LI^s\rho(s) ds \nonumber\\
			& & +\eta_\rho(t) + i\int_0^te^{-i(t-s)\Lb}\LI^s\eta_\rho(s) ds \nonumber\\
			& & + \mathcal{O}(h_\opB^2).\label{eq:app:formalstate}
	\end{eqnarray}

From the expression above, one can calculate the average of the observable $\opA$ by applying the Schr\"odinger-Heisenberg equivalence and select $\rho(0)=\equi$, one obtains
	\begin{eqnarray}
		\aveb{\opA(t)}_\rho &=& \avef{\opA} + i\tr\left( \int_0^t e^{-i(t-s)\Lb}\LI^s\equi\opA ds\right)\nonumber\\
		& &+\tr\left[\eta_{\equi}(t)\opA + i\int_0^te^{-i(t-s)\Lb}\LI^s\eta_{\equi}(s)\opA ds  \right]\nonumber\\
			& & + \mathcal{O}(h_\opB^2).
		\label{eq:app:Aaveragedynamic}
	\end{eqnarray}
Next consider the integrand in the second term on right hand side of the equation, one can see that 
	\begin{widetext}
	\begin{eqnarray}
	\label{eq:app:Sch2Hei4non}
		\tr\left(e^{-i(t-s)\Lb}\left[\opB,\equi\right]\opA\right)  &=& \tr\left(\opUb(t)\opUb^\dagger(s)\left[\opB,\equi\right]\opUb(s)\opUb^\dagger(t)\opA\right)= \tr\left(\equi\left[\opA(t),\opB(s)\right]\right),
	\end{eqnarray}
	\end{widetext}
where $\opUb(t) = e^{-it\opHb}$, $\opA(t) = \opUb^\dagger(t)\opA\opUb(t)$ and $\opB(t) = \opUb^\dagger(t)\opB\opUb(t)$. The relations above arise from the cyclic invariance of trace operator and the fact that $e^{-it\Lb}\mathbf{X} = e^{-itH}\mathbf{X}e^{itH}$.

The linear response function is defined by
	\begin{equation}
	\resp{\opA}{\opB}(t,t') = \dfrac{\partial \aveb{\opA(t)}_\rho}{\partial \h(t')}\bigg\vert_{\h=0}.
	\label{eq:app:response}
	\end{equation}	
Then we obtain, for $0\leq t'\leq t$,
	\begin{equation}
		\resp{\opA}{\opB}(t,t') = 2i\theta(t-t')\{\tr\left(\equi\left[\opA(t),\opB(t')\right]\right) + \Delta_{\opA\opB}(t,t') \},
	\end{equation}	
	where
	\begin{widetext}
	\begin{equation}
	\label{eq:app:extraterm}
	\Delta_{\opA\opB}(t,t') := \dfrac{1}{2}\tr\left[\left(\int_0^{t'} e^{-i(t-t')\Lb}\left[\opB,e^{-i(t'-s)\Lb} \covdiff{s}{\opHb}{\rho} \right] ds\right)\opA\right].
	\end{equation}
	\end{widetext}
In physical sense, the term $\tr\left(\equi\left[\opA(t),\opB(t')\right]\right)$ can be related to an symmetric correlation \cite{2011Cugliandolo}; where correlation is usually defined as a two-epoch time correlation function between operators $\opA$ and $\opB$ in a given state. In this case:
	\begin{equation}
		\corr{\opA}{\opB}\left(t,t'\right) := \avef{\opA(t)\opB(t')}.
	\end{equation}
	One can then define symmetric and anti-symmetric correlations respectively by
	\begin{eqnarray}
		\corr{\opA}{\opB}^{-}\left(t,t'\right) &:=& \avef{ \left[\opA(t),\opB(t')\right]}, \\
		\corr{\opA}{\opB}^{+}\left(t,t'\right) &:=& \avef{ \{\opA(t),\opB(t')\} },
	\end{eqnarray}
where $\{\opA,\opB\}:= \opA\opB + \opB\opA$. Thus, the modified form of the Kubo's formula can be written as
	\begin{equation}
	\label{eq:app:Kubo}
		\resp{\opA}{\opB}(t,t') = 2i\theta(t-t')\left[\corr{\opA}{\opB}^{-}\left(t,t'\right) + \Delta_{\opA\opB}(t,t') \right],
	\end{equation}
where $\theta(t)$ is a Heaviside function defined by $\theta(t)=1$ for $t\geq 0$ and $\theta(t)=0$ for $t<0$. Now consider the conditions to obtain the original FDR from Eq.\eqref{eq:app:Kubo}. First, it follows that   
	\begin{align}
	\corr{\opA}{\opB}\left(t,t'\right) &= \tr\left(\equi\opA(t)\opB(t')\right)\nonumber\\
	 &= \tr\left(\equi\opA(t-t')\opB\right) \nonumber\\
	 &= \corr{\opA}{\opB}\left(t-t',0\right)
	\end{align}
by the cyclic invariance of trace. Thus, 
	\begin{equation}
	\corr{\opA}{\opB}^{\pm}\left(t,t'\right) = \corr{\opA}{\opB}^{\pm}\left(t-t',0\right).
	\end{equation}
Using the short-hand writing $(t,t')\rightarrow(t)$ when $t'=0$, so that $\corr{\opA}{\opB}\left(t,t'\right) \rightarrow \corr{\opA}{\opB}\left(t\right)$, 
we obtain
	\begin{eqnarray}
		\corr{\opA}{\opB}\left(t\right)	&=&	\tr\left(\equi\opA(t)\opB \right)\nonumber\\
				&=&	\dfrac{\tr\left(\opUb(t)\opB\opUb^\dagger(i\beta)\opUb^\dagger(t)\opA \right)}{\tr\opUb(i\beta)}\nonumber\\
				&=&	\tr\left(\equi\opUb^\dagger(-t)\opUb^\dagger(-i\beta)\opB\opUb(-i\beta)\opUb(-t)\opA \right)\nonumber\\
		 \corr{\opA}{\opB}\left(t\right)	&=& \corr{\opB}{\opA}\left(-t-i\beta\right).
		\label{eq:app:commutincorr}
	\end{eqnarray}
One can see that the relation above is another version of the Kubo-Martin-Schwinger(KMS) condition, and the evolution along imaginary time arises here.

Suppose the interaction of the composite system with the bath begins at $t'= 0$ (or equivalently, the probe start at time $t' = 0$ when the system is still in adjusted equilibrium state.) Then one can verify that $\Delta_{\opA\opB}(t) = 0$; and Eq.\eqref{eq:app:Kubo} becomes
	\begin{equation}
	\resp{\opA}{\opB}(t) = 2i\theta(t)\corr{\opA}{\opB}^{-}(t).
	\label{eq:app:bFDR}
	\end{equation}
It would be mentioned here that the resulting equation can be approached by another reason or assumption, for instance, when one forces the system so that $\rho_\omega(t)$ always lies on its mean state $\rho(t)$ -- which is an alternative expression for \emph{mean-field approximation} also.

For an integrable function $g(t)$, we define its Fourier transform as
	\begin{equation*}
	 \hat{g}(\lambda) \equiv \mathcal{F}\left( g(t) \right) := \dfrac{1}{\sqrt{2\pi}}\int_{-\infty}^\infty e^{-it\lambda}g(t)\, dt, 
	\end{equation*}
	where $\lambda$ is a complex number.
Taking the transform of Eq.\eqref{eq:app:bFDR}, so
	\begin{eqnarray}
		\hresp{\opA}{\opB}(\lambda) &=& \hcorr{\opA}{\opB}^{-}(\lambda)\nonumber\\
			&=&	\left(1-e^{-\beta\lambda} \right)\hcorr{\opA}{\opB}(\lambda).
	\end{eqnarray}
The last relation is a direct consequence of Eq.\eqref{eq:app:commutincorr}. Furthermore, one finds that  
	\begin{equation}
		\hcorr{\opA}{\opB}^{+}(\lambda) = \left(1+e^{-\beta\lambda} \right)\hcorr{\opA}{\opB}(\lambda).
	\end{equation}		
Combine terms together in the equations above, one can obtain  the standard fluctuation-dissipation relation
	\begin{equation}
		\hresp{\opA}{\opB}(\lambda) = \tanh\left(\dfrac{\beta\lambda}{2}\right)\hcorr{\opA}{\opB}^{+}(\lambda).
	\label{eq:app:FDR}
	\end{equation}

\end{document}